\documentclass[a4paper,12pt]{article}

\usepackage[a4paper,margin=1in]{geometry}
\usepackage{amsmath,amsthm,amssymb}
\usepackage{natbib}
\usepackage{graphicx}
\usepackage{array}
\usepackage{setspace}
\usepackage[dvipsnames]{xcolor}
\usepackage{subcaption}
\usepackage{booktabs}
\usepackage{comment}
\usepackage{enumitem}
\usepackage{algorithm}
\usepackage{algorithmic}
\usepackage{threeparttable}
\usepackage{hyperref}
\usepackage{rotating}
\usepackage{siunitx}
\usepackage{comment}

\bibpunct{(}{)}{;}{a}{,}{,}

\numberwithin{equation}{section}

\newtheorem{theorem}{Theorem}

\newtheorem{proposition}{Proposition}

\theoremstyle{remark}

\theoremstyle{definition}

\newcommand{\csection}[1]{
  \phantomsection
  \refstepcounter{section}
  \begin{center}
    {\bfseries\large\arabic{section}. #1}
  \end{center}
}

\newcommand{\csubsection}[1]{
  \phantomsection
  \refstepcounter{subsection}
  \begin{center}
    {\itshape\arabic{section}.\arabic{subsection}. #1}
  \end{center}
}

\newcommand{\E}{\mathrm{E}}
\newcommand{\mR}{\mathbb{R}}

\newcommand{\F}{\mathrm{F}}

\title{\bfseries Quantile Social Autoregressive Model}

\author{%
Liyuan Wang$^{1}$,
Danyang Huang$^{1}$%
\thanks{Correspondence to: Danyang Huang, Center for Applied Statistics and
School of Statistics, Renmin University of China, No.~59 Zhongguancun Street,
Haidian District, Beijing 100872, China; Email: dyhuang@ruc.edu.cn.
Wei Lan, Center of Statistical Research, Southwestern University of Finance
and Economics, 555 Liutai Avenue, Wenjiang District, Chengdu, Sichuan 611130,
China; Email: \mbox{lanwei@swufe.edu.cn}.},
Wei Lan$^{2}$\footnotemark[1],
and Chih-Ling Tsai$^{3}$\\[0.8em]
\parbox{0.95\textwidth}{\centering\normalsize
$^{1}$ Center for Applied Statistics and School of Statistics,
Renmin University of China, Beijing, China\\[0.25em]
$^{2}$ Center of Statistical Research,
Southwestern University of Finance and Economics, Chengdu, China\\[0.25em]
$^{3}$ Graduate School of Management,
University of California, Davis, CA, USA
}}

\date{}

\begin{document}

\maketitle

\begin{singlespace}
\begin{abstract}
Research on modelling peer effects has predominantly relied on linear-in-means models. However, averaging peer responses prevents these models from capturing the effects of extreme peer behavior. To address this limitation, we
propose a quantile social autoregressive model based on a novel quantile social
norm, which uses empirical quantiles of peer responses. By treating the quantile
level as an unknown parameter estimated directly from the data, our approach
identifies which segment of the peer response distribution most strongly influences individual behavior. To estimate the model, we introduce new moment conditions using pseudo-response instruments. Because the quantile social norm is nonsmooth, we apply kernel smoothing to the instruments and residuals, ensuring valid statistical inference. Additionally, we establish equilibrium existence and uniqueness and derive the identification conditions. Furthermore, we prove the consistency and asymptotic normality of our proposed estimator. Finally, Monte Carlo experiments examine the finite-sample performance of the estimator, and an empirical application illustrates the interpretation of the estimated quantile levels and their corresponding peer effects.

\end{abstract}
\end{singlespace}

\newpage

\csection{INTRODUCTION}

\csubsection{Background and Related Literature}

Network data consist of nodes, nodal attributes, and edges that encode relationships between nodes. Individual responses can be influenced not only by personal covariates but also by the responses of connected neighbors--a phenomenon known as a peer effect \citep{Bramoulle2009, LeeLiuLin2010}. Researchers often employ the linear-in-means (LIM) model to characterize these interactions; see \citep{Manski1993, Bramoulle2009, blume2015linear, BasseDingFellerToulis2024}. This model assumes that an individual's response depends linearly on the average responses of their peers\footnote{This formulation is conceptually and mathematically aligned with the structure of spatial autoregressive (SAR) models \citep{Ord1975, LeeLiuLin2010, Bramoulle2020}, which are widely applied in network settings to capture social dependence.}. Furthermore, this framework supports extensive empirical research across diverse domains. These include education and workplace performance \citep{Sacerdote2001, Carrell2009}, criminal behavior \citep{Glaeser1996, Bayer2009}, investment decisions \citep{Bursztyn2014}, and digital entertainment consumption \citep{Jo2025}.

To characterize social dependence, the social norm specifies how an individual's response depends on their neighbors' collective behaviour \citep{ushchev2020social, EconometricaPeerEffects}. The standard LIM framework assumes that the social norm corresponds to the arithmetic mean of peer responses. However, this restriction prevents the model from capturing the diverse peer effects observed in real-world networks \citep{LeLi2022}. Furthermore, empirical research shows that using distributional information (such as quantiles) or rank information provides deeper insights than relying solely on the mean \citep{Chernozhukov2024, Lin2024, ZhuEtAl2019NQAR, Denning2023Rank}. For example, peer effects often depend on specific segments of a group: basketball players may be influenced most by top-tier teammates \citep{Tao2014}, whereas students' academic performance can suffer from low-achieving or disruptive peers \citep{Lazear2001}. Evidence from other domains also suggests that individuals may respond to specific peer subgroups rather than to the group average \citep{Bietenbeck2020, Hahn2020, Diaz2021, denton2026conformity}. Because the mean fails to capture these nuanced dynamics, we adopt a framework that incorporates quantile- and rank-based peer information.

To identify peer effects arising from different influential groups with a potentially nonlinear social norm, \citet{EconometricaPeerEffects} proposed a model based on a constant elasticity of substitution (CES) social norm (hereafter referred to as the CES model). By varying the elasticity parameter, this framework flexibly captures peer effects from different segments of the peer distribution. Specifically, the CES social norm can incorporate various types of neighbour information to model social influence, such as the mean, maximum, or minimum of the neighbours' responses. Notably, the LIM model is a special case of this general framework when the elasticity parameter equals 1. \citet{wang2023generalizing} employed a similar idea in the context of graph neural networks by introducing nonlinear aggregation functions that allow adaptive transitions between average and extreme peer influences.

Despite its flexibility, the CES model has several limitations regarding interpretability and applicability \citep{EconometricaPeerEffects}. First, the elasticity parameter spans a wide range, which makes reliable interpretation difficult in practice. Second, the norm is highly sensitive to extreme values. When the elasticity parameter is large in magnitude, the norm concentrates on a small subset of neighbours, mimicking a max- or min-type operator. Consequently, an outlier peer value can dominate the aggregated effect. Third, it requires all response variables to be strictly positive, which limits its applicability when outcomes are zero or negative.

A possible approach to mitigating the limitations of the CES model is to use peer ranks or quantile information to explore peer effects \citep{Herstad2024,Houndetoungan2025}. In the existing literature, however, network models involving quantiles focus primarily on modelling conditional quantiles of individual responses rather than quantiles of peer responses. For example, \citet{ZhuEtAl2019NQAR} introduced a network quantile autoregression model that allows the effects of lagged responses from connected nodes and node-specific characteristics to differ across quantile levels. Subsequently, \citet{Xu2024} extended this framework by incorporating contemporaneous network effects and observed common factors with quantile-specific coefficients. Conditional quantile models have also been developed for spatially dependent data. For example, \citet{Ando2023} proposed a spatial panel quantile model with unobserved heterogeneity, while \citet{ChenTokdar2021} developed a joint quantile regression framework that accounts for spatial dependence through a copula process. Additionally, \citet{XieLi2018} used quantile-based measures of dependence to infer high-dimensional network structures.

Existing approaches that use peer ranks or quantile information to study peer effects typically require either bounded nodal degrees \citep{Herstad2024} or pre-selected quantile levels \citep{Houndetoungan2025}. To bridge this gap, we develop an innovative model for estimating latent quantile levels, which allows node degrees to grow alongside the network scale. Our main contributions are listed below.

\csubsection{Novelty, Challenges, and Contributions}

To address the limitations of existing peer-response aggregation methods, we propose a quantile social autoregressive (QSAR) model in which the $z$-th quantile of the neighbours' responses serves as the social norm. This specification is novel and offers three advantages: (a) an intuitive interpretation of the influential peer segment; (b) resistance to extreme peer responses; and (c) a simpler parameter search because $z\in(0,1)$, compared with the unbounded elasticity parameter in \citet{EconometricaPeerEffects}. Unlike models with pre-selected quantile levels, the QSAR model estimates $z$ directly from the data. This provides a data-driven approach to identify which segment of the peer response distribution drives individual behaviour.

Despite the novelty of our proposed model, we encounter two major challenges in its theoretical analysis, estimation, and practical implementation.
One challenge arises from the absence of a closed-form reduced representation and the endogeneity of the peer-response aggregate. We establish the existence and uniqueness of the equilibrium and construct oracle instruments from pseudo responses to obtain moment conditions for identification. Replacing the oracle instruments with feasible sample-based counterparts yields a generalized method of moments (GMM) estimator. We show that the discrepancy between the feasible and oracle sample moments is asymptotically negligible.

The other challenge arises from the nondifferentiability of the quantile social norm and its stronger local dependence relative to the LIM model. We apply kernel smoothing to the instruments for feasible estimation and to the quantile social norm in the residual for statistical inference. This produces a doubly smoothed GMM estimator. Under appropriate bandwidth and network conditions, we establish the consistency of both GMM estimators and the asymptotic normality of the doubly smoothed GMM estimator.

In summary, our contributions not only open a new avenue for studying peer effects in social networks, but also expand quantile applications through novel estimation methods, algorithms, and theory.

\csubsection{Organization}

The remainder of the paper is organized as follows. Section 2 defines the quantile social norm and introduces the QSAR model. Section 3 constructs the moment conditions and outlines the GMM estimation algorithm. Section 4 analyzes model identification and equilibrium and establishes the asymptotic properties of the estimators. Finally, Section 5 evaluates finite-sample performance using Monte Carlo experiments and illustrates the estimated peer effects through a real-world application.

\csection{Quantile Social Autoregressive Model}

\csubsection{Notation and Linear-in-Means Model with CES Norm}

We begin by introducing the notation used throughout the paper. Consider a network with $n$ nodes. Let $\boldsymbol{A}=(a_{i j})\in\{0,1\}^{n\times n}$ be the adjacency matrix, where $a_{i j}=1$ indicates a directed edge from node $i$ to node $j$ ($i \neq j$) and $a_{i j}=0$ otherwise. For completeness, set $a_{ii}=0$ for $1\leqslant i \leqslant n$. Define the row-normalized weight matrix $\boldsymbol{W}=(w_{i j})\in\mathbb{R}^{n\times n}$ with $w_{i j}=a_{i j}/d_i$, where $d_i=\sum_{j=1}^n a_{i j}$ is the out-degree of node $i$.
For each node $i$, let $\mathcal{A}_i=\{j: a_{ij}=1\}$ denote its set of neighbors, $Y_i$ represent the response, and $X_i \in \mathbb{R}^p$  be the exogenous covariate vector of fixed dimension $p$. We then stack the responses and covariates into the response vector $\boldsymbol{Y}=(Y_1, \cdots, Y_n)^\top \in \mathbb{R}^{n}$ and the covariate matrix $\boldsymbol X=(X_1,\cdots,X_n)^\top\in\mathbb{R}^{n\times p}$, with $\boldsymbol X$ and $\boldsymbol A$ treated as fixed throughout the theoretical analysis.
In this paper, we consider several standard norms. For a random variable $u$, the sub-Gaussian norm is defined as $\|u\|_{\psi_2}=\inf\{K>0:\E\{\exp(u^2/K^2)\}\leq 2\}$ \citep{Vershynin2018}. For vectors, we use the \(\ell _{2}\)- and \(\ell _{\infty }\)-norms. For matrices, we employ the \(\ell _{1}\)-norm ($\|\cdot\|_1$) and the Frobenius norm ($\|\cdot\|_{\F}$).

We next review the LIM model with CES social norms \citep{EconometricaPeerEffects}. The model is formulated as follows:
\begin{equation}
\label{eq:CES}
Y_i=\lambda\widetilde Y_{-i}(\gamma) + X_i^{\top}\boldsymbol{\beta} + \varepsilon_i,
\end{equation}
where
$$\widetilde Y_{-i}(\gamma)=\left(\sum_{j=1}^{n} w_{i j} Y_{j}^{\gamma}\right)^{1/\gamma}$$
is the CES social norm of node $i$'s neighbours, $\boldsymbol{\beta}$ denotes the covariate coefficients, $\lambda$ captures the intensity of peer effects, and $\gamma\in[-\infty,+\infty]$ is the elasticity parameter. The CES social norm is well defined when the responses are strictly positive. When $\gamma=1$, it reduces to the classical LIM model \citep{Manski1993}.
In addition, $\widetilde Y_{-i}(\gamma)$ converges to $\max_{j\in\mathcal{A}_i}Y_{j}$ as $\gamma\to+\infty$ and to $\min_{j\in\mathcal{A}_i}Y_{j}$ as $\gamma\to-\infty$.
Despite its flexibility, the elasticity parameter $\gamma$ lacks a clear economic interpretation. Because a specific value of $\gamma$ does not reveal which segment of the peer distribution drives individual behaviour, the empirical results remain difficult to interpret.

\csubsection{Quantile Social Norm and Quantile Social Autoregressive Model}

To address the limitations outlined above, we propose adopting quantile social norms in place of the CES social norm. For each node $i$, let $Y_{i,(1)}\leqslant \cdots \leqslant Y_{i,(d_i)}$ denote the order statistics of $\{Y_j:j\in\mathcal A_i\}$. For a given quantile level $z\in(0,1)$ and $d_i\geqslant 2$, define the {\it quantile social norm} in  a linearly interpolated form as follows:
\begin{equation}
\label{eq:quantilenorm}
Y_{i}^{\dagger}(z) =\{1-\alpha_i(z)\}Y_{i,(k_i(z))} +\alpha_i(z)Y_{i,(k_i(z)+1)},
\end{equation}
where $\alpha_i(z)=\ell_i(z)-k_i(z)$, $\ell_i(z)=1+(d_i-1)z$, and $k_i(z)=\lfloor\ell_i(z)\rfloor$ is the greatest integer not exceeding $\ell_i(z)$.
The classical empirical quantile is discontinuous and remains unchanged over a range of quantile levels. However, linear interpolation between the interpolation knots $t_{ik}=(k-1)/(d_i-1)$, for $k=1,\cdots,d_i$, allows the quantile social norm to vary continuously with $z$.
When $z=0.5$, $Y_{i}^{\dagger}(z)$ represents the median of peer responses. In addition, as $z\to 0^+$ and $z\to 1^-$, $Y_{i}^{\dagger}(z)$ approaches the minimum and maximum of $\{Y_j:j\in\mathcal A_i\}$, respectively, which aligns with the $\gamma\to-\infty$ and $\gamma\to+\infty$ cases of the CES norm in~\eqref{eq:CES}.
Unlike the CES elasticity parameter $\gamma$, the quantile level $z\in(0,1)$ directly indicates which segment of the peer response distribution is most strongly associated with an individual's response.
Accordingly, we define the quantile social autoregressive model (QSAR) as follows:
\begin{equation}
\label{eq:model}
Y_i = \lambda Y_{i}^{\dagger}(z) + X_i^\top \boldsymbol \beta + \varepsilon_i, \quad i=1,\cdots,n,
\end{equation}
where $Y_{i}^{\dagger}(z)$ is defined in~\eqref{eq:quantilenorm}, $\lambda$ measures the strength of the peer effect at the quantile level $z$, and the errors $\{\varepsilon_i\}_{i=1}^n$ are independent random variables satisfying $E(\varepsilon_i)=0$ and $\operatorname{Var}(\varepsilon_i)<\infty$ for each $i$.

To facilitate the estimation and theoretical analysis, we express the model in vector form. Let $\boldsymbol Y^{\dagger}(z)=\bigl(Y_{1}^{\dagger}(z),\cdots,Y_{n}^{\dagger}(z)\bigr)^\top$
and
$\boldsymbol\varepsilon=(\varepsilon_1,\cdots,\varepsilon_n)^\top$.
Then the QSAR model can be written as follows:
\begin{equation}
\label{eq:model_matrix}
\boldsymbol Y
=
\lambda\boldsymbol Y^{\dagger}(z)
+
\boldsymbol X\boldsymbol\beta
+
\boldsymbol\varepsilon.
\end{equation}
The QSAR model in \eqref{eq:model_matrix} provides several key advantages. First, the parameter $z$ has an intuitive interpretation, specifying which segment of the peer response distribution is most closely associated with individual behaviour. Second, unlike classical empirical quantiles, the interpolation avoids flat regions so that even small changes in $z$ affect the peer norm. Third, the quantile-based norm resists outliers and heavy-tailed peer responses by relying on local order rather than the full average. Furthermore, the quantile social norm is more robust to noisy links than the neighbour average.

\begin{figure}[htbp]
  \centering

  \includegraphics[width=\textwidth]{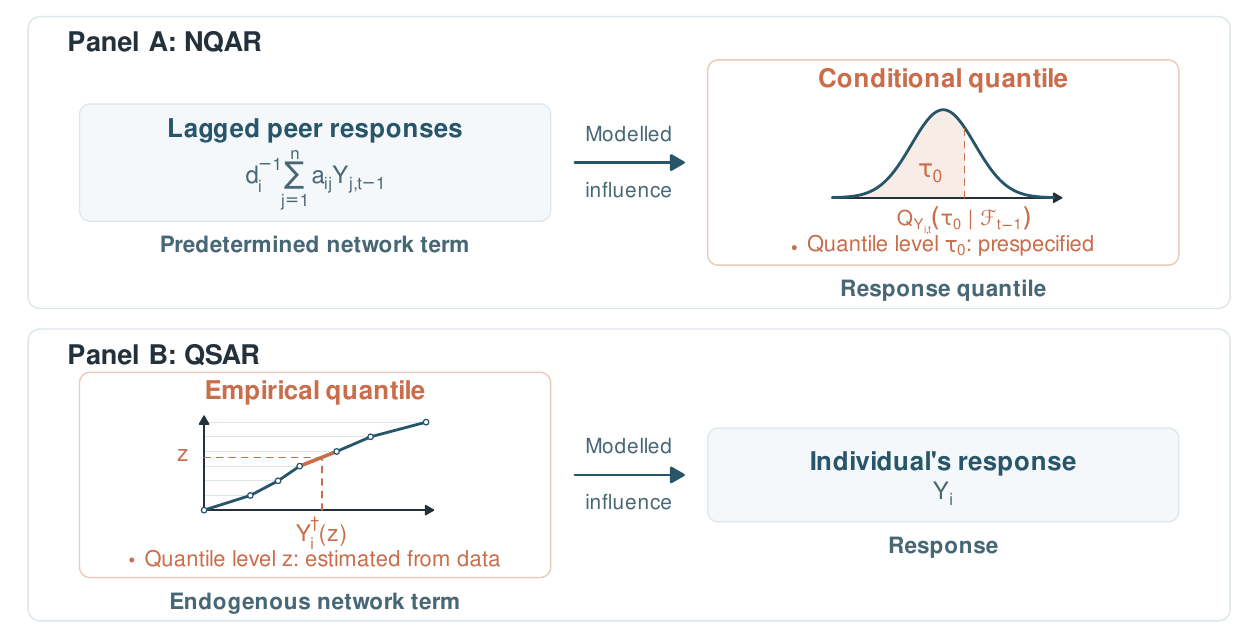}

  \caption{Comparison between NQAR and QSAR. Panel A illustrates how NQAR models the conditional quantile $Q_{Y_{i,t}}(\tau_0\mid\mathcal F_{t-1})$ of node $i$'s response at the specified quantile level $\tau_0$. Panel B depicts how QSAR captures peer effects through the quantile social norm $Y_i^{\dagger}(z)$ of node $i$'s neighbours. The detailed NQAR model structure is presented in Appendix S7.}

  \label{fig:LIM_vs_QSAR}
\end{figure}

\noindent\textbf{Remark: (Comparison with Network Quantile Autoregression)}
It is critical to address the two main differences between our proposed QSAR model and the Network Quantile Autoregression (NQAR) model
introduced by \citet{ZhuEtAl2019NQAR}.
First, the NQAR model characterizes network dependence through lagged peer responses $d_i^{-1} \sum_{j} a_{ij} Y_{j,t-1}$.
Hence, their spatial regressor is measurable with respect to the
$\sigma$-algebra $\mathcal F_{t-1}$ generated by information up to time
$t-1$, thereby avoiding the endogeneity inherent in simultaneous
spatial interaction models. In contrast, QSAR focuses on capturing contemporaneous peer effects.
Second, the NQAR model considers an individual's theoretical conditional quantile, whereas our model employs the empirical quantile of neighbours' responses; Figure \ref{fig:LIM_vs_QSAR} illustrates these differences. This fundamental difference also appears in existing quantile autoregression models \citep{koenker2006quantile}.

Despite its appealing properties, the QSAR model poses two key challenges for theoretical analysis. First, the linear-in-means (LIM) model yields an explicit reduced-form representation $\boldsymbol Y=(\boldsymbol I_n-\lambda\boldsymbol W)^{-1}(\boldsymbol X\boldsymbol\beta+\boldsymbol\varepsilon)$ \citep{Ord1975,anselin2013spatial}, where $\boldsymbol I_n\in\mathbb R^{n\times n}$ is the $n\times n$ identity matrix. However, the QSAR model lacks any comparable closed-form representation. Second, while the quantile social norm $Y_i^{\dagger}(z)$ is continuous in $z$, it can be nondifferentiable at the interpolation knots. Section 4 tackles these two challenging tasks by proving equilibrium existence and uniqueness without a reduced form and by smoothing the quantile social norm for valid asymptotic theory.

Before presenting the estimation procedure, we introduce the parameter space and its associated conditions. Let the parameter vector be $\boldsymbol{\theta}=(\boldsymbol{\beta}^\top,\lambda,z)^\top\in\mathbb R^{p+2}$, with the true value $\boldsymbol{\theta}_0=(\boldsymbol{\beta}_0^\top,\lambda_0,z_0)^\top$. We define the parameter space as $\boldsymbol{\Theta}=\mathcal B\times\Lambda\times\mathcal Z$, where $\mathcal B\subset\mathbb R^p$ is compact, $\Lambda=[-1+c_\lambda,-c_\lambda]\cup[c_\lambda,1-c_\lambda]$, and $\mathcal Z=[\delta,1-\delta]$ for some small $c_\lambda\in(0,1/2)$ and $\delta\in(0,1/2)$.

Note that bounding the quantile parameter $z_0\in\mathcal Z$ away from $0$ and $1$ provides a dual benefit: it strengthens robustness against outliers and maintains theoretical tractability by avoiding the ill-behaved asymptotic variance associated with extreme quantiles \citep{mosteller1946some,chernozhukov2005extramal}. In addition, the domain of the peer-effect coefficient, $\Lambda$, excludes a neighbourhood of zero to ensure identification. Indeed, if $\lambda_0=0$, the quantile-social-norm term in Equation \eqref{eq:model_matrix} vanishes, eliminating the model's dependence on $z_0$ and rendering $z_0$ unidentifiable. We therefore conduct a preliminary test of $\mathrm H_0:\lambda_0=0$ against $\mathrm H_1:\lambda_0\neq0$ before estimating the model. This test is constructed by aggregating fixed-$z$ statistics across a prespecified grid of candidate quantile levels. Section S4 of the supplementary material provides the detailed implementation procedure. Furthermore, the condition $|\lambda_0|<1$ (see also Section 4.1) ensures a unique and stable equilibrium. This mirrors the standard LIM stability restriction used to ensure the invertibility of $(\boldsymbol I_n-\lambda_0\boldsymbol W)$. To satisfy these constraints, we define $\Lambda$ as a union of closed intervals, guaranteeing the compact parameter space needed for our theoretical derivations.

\csection{Moment Conditions and Estimation Algorithms}

\csubsection{Instrument Construction and Moment Conditions}

We estimate the parameter vector $\boldsymbol{\theta}$ of model \eqref{eq:model_matrix} using a generalized method of moments (GMM) framework with appropriate moment conditions. We first consider the ordinary least squares (OLS) estimator,
\begin{equation}
\label{eq:OLS}
\widehat{\boldsymbol\beta}=(\boldsymbol X^\top\boldsymbol X)^{-1}\boldsymbol X^\top\boldsymbol Y
=\boldsymbol\beta_0+\lambda_0(\boldsymbol X^\top\boldsymbol X)^{-1}\boldsymbol X^\top\boldsymbol Y^{\dagger}(z_0)+(\boldsymbol X^\top\boldsymbol X)^{-1}\boldsymbol X^\top\boldsymbol\varepsilon,
\end{equation}
where
 $\boldsymbol Y$ is generated under model (\ref{eq:model_matrix}) at the true parameter $\boldsymbol{\theta}_0=(\boldsymbol{\beta}_0^\top,\lambda_0,z_0)^\top$.
Given $\E(\boldsymbol\varepsilon)=0$,
its population counterpart is
$\boldsymbol\beta^*=\boldsymbol\beta_0+\lambda_0Q_X^{-1}q_{XY}$, where  $Q_X=n^{-1}\boldsymbol X^\top\boldsymbol X$ and $q_{XY}=n^{-1}\boldsymbol X^\top\E\{\boldsymbol Y^{\dagger}(z_0)\}$.
Since $\boldsymbol\beta^*$ is a non-stochastic population coefficient whereas $\widehat{\boldsymbol\beta}$ depends on $\boldsymbol\varepsilon$, we use $\boldsymbol\beta^*$ to define the oracle pseudo-responses $\widetilde Y_i^0=X_i^\top\boldsymbol\beta^*$ and stack them into the vector $\widetilde{\boldsymbol Y}^0=(\widetilde Y_1^0,\cdots,\widetilde Y_n^0)^\top$.
Consequently, $\widetilde{\boldsymbol Y}^0$ is independent
of $\boldsymbol\varepsilon$, allowing us to use it to construct oracle moment conditions to identify unknown parameters.
We then introduce the following two moment conditions, with (M2) corresponding to the oracle moment condition.
\begin{itemize}
\item[(M1)] $\E(\boldsymbol X^\top\boldsymbol\varepsilon)=\mathbf{0}_{p\times 1}$, where $\mathbf{0}_{p\times 1}$ is a $p$-dimensional vector of zeros;
\item[(M2)] $\E[\lambda_0\{\widetilde{\boldsymbol Y}^{\dagger,0}(z_0)\}^\top\boldsymbol\varepsilon]=0$, where the $i$-th element of $\widetilde{\boldsymbol Y}^{\dagger,0}(z)$ is obtained from Equation \eqref{eq:quantilenorm} by replacing each neighbour response $Y_j$ with $\widetilde Y_j^0$.
\end{itemize}

In practice, $\boldsymbol\beta^*$ is unknown so we replace it with its sample estimator $\widehat{\boldsymbol\beta}$.
We then define the feasible pseudo responses $\widetilde Y_i=X_i^\top\widehat{\boldsymbol\beta}$ and stack them into the vector $\widetilde{\boldsymbol Y}=(\widetilde Y_1,\cdots,\widetilde Y_n)^\top$. Consequently, the feasible instruments constructed from $\widetilde{\boldsymbol Y}$  serve as  sample-based approximations to the oracle instruments. Furthermore,
the second feasible moment condition shares the same form as (M2), except that
$\widetilde{\boldsymbol Y}^{\dagger,0}(z)$ is replaced by $\widetilde{\boldsymbol Y}^{\dagger}(z)$.
Note that this feasible moment condition is adopted from \cite{EconometricaPeerEffects}.

Since $\widehat{\boldsymbol\beta}$ in ({\ref{eq:OLS}}) is correlated with $\boldsymbol\varepsilon$ via $\boldsymbol Y$,  the feasible instruments computed from $\boldsymbol Y$ can depend on $\boldsymbol\varepsilon$. Consequently,  the expectation of the feasible version of (M2)  is generally non-zero at $\boldsymbol\theta_0$. To bridge this theoretical gap, we show that the discrepancy between the feasible and oracle sample moments is asymptotically negligible.

\csubsection{Kernel Smoothing Condition and Estimation Algorithms}

Although the oracle moment condition (M2) holds, its instrument $\widetilde{\boldsymbol Y}^{\dagger,0}(z)$ is not differentiable at the node-specific interpolation knots $\{t_{ik}\}_{k=1}^{d_i}$. To address this issue, we apply convolution smoothing, a kernel smoothing approach widely used in quantile estimation \citep{KaiLiZou2010,FernandesGuerreHorta2021,TanWangZhou2022}, to ensure the oracle instrument is differentiable.
To construct the convolution smoothing function, we first define a constant-extended version $\widetilde Y_{i,\mathrm{ext}}^{\dagger,0}(v)$ of $\widetilde Y_i^{\dagger,0}(\cdot)$.
This extended function coincides with $\widetilde Y_i^{\dagger,0}(\cdot)$ over the interior domain $0<v<1$; for $v\leqslant0$, it is fixed to $\min_{j\in\mathcal A_i}\widetilde Y_j^0$, while for $v\geqslant1$, it takes the constant value $\max_{j\in\mathcal A_i}\widetilde Y_j^0$. We then apply kernel smoothing to $\widetilde Y_{i,\mathrm{ext}}^{\dagger,0}(v)$ using the kernel $k(\cdot)$ together with node-dependent smoothing bandwidths $\{\tau_i\}_{i=1}^n$, which gives the smoothed function
$\widetilde Y_i^{*,0}(z)=\int k(u)\widetilde Y_{i,\mathrm{ext}}^{\dagger,0}(z-\tau_i u)\,\mathrm{d}u.$
Finally, we stack these smoothed functions to form the vector
$\widetilde{\boldsymbol Y}^{*,0}(z)=\big(\widetilde Y_1^{*,0}(z),\,\cdots,\,\widetilde Y_n^{*,0}(z)\big)^\top.$
Based on this smoothed vector, we substitute the original condition (M2) with the smoothed oracle moment condition stated below:
\begin{itemize}
\item[(M2$^*$)] $\E[\lambda_0\{\widetilde{\boldsymbol Y}^{*,0}(z_0)\}^{\top}\boldsymbol\varepsilon]=0$.
\end{itemize}
Because $\widetilde{\boldsymbol Y}^{*,0}(z_0)$ is constructed from the oracle pseudo-responses and is independent of $\boldsymbol\varepsilon$, the validity of this moment condition remains unaffected by the smoothing.

In practice, $\widetilde{\boldsymbol Y}^{*,0}(z_0)$ is unknown. We therefore define the feasible smoothed instrument as $\widetilde Y_i^*(z)=\int k(u)\widetilde Y_{i,\mathrm{ext}}^\dagger(z-\tau_i u)\,du$, where $\widetilde Y_{i,\mathrm{ext}}^\dagger$ refers to the constant-extended function constructed analogously from $\widetilde Y_i^\dagger$.
Stacking these elements into the vector $\widetilde{\boldsymbol Y}^*(z)$, we derive the feasible version of (M2$^*$) by replacing $\widetilde{\boldsymbol Y}^{*,0}(z)$ with $\widetilde{\boldsymbol Y}^*(z)$. Under these extensions, both the oracle and feasible smoothed instruments are twice differentiable for all $z\in\mathcal Z$. This regularity property is critical for implementing Taylor expansions of smoothed sample moments when conducting the asymptotic analysis.

For any fixed $z$, moment conditions (M1) and (M2$^*$) are sufficient for us to estimate $\boldsymbol\beta$ and $\lambda$ via GMM. However, they are insufficient to identify the true quantile level $z_0$.
Hence, inspired by \cite{EconometricaPeerEffects}, we  construct a third moment condition using the derivative of the smoothed peer influence with respect to $z$:
\begin{itemize}
\item[(M3)] $\E[\lambda_0\{\partial\widetilde{\boldsymbol Y}^{*,0}(z_0)/\partial z\}^{\top}\boldsymbol\varepsilon]=0$.
\end{itemize}
We then construct the feasible moment condition for (M3) by replacing the oracle derivative $\partial\widetilde{\boldsymbol Y}^{*,0}(z)/\partial z$ with its feasible counterpart, $\partial\widetilde{\boldsymbol Y}^{*}(z)/\partial z$.

Using parameter identification through moment conditions (M1), (M2$^*$) and (M3) and their feasible counterparts, we employ GMM and control the feasible-oracle discrepancy in the sample moments to establish the asymptotic properties of the resulting estimators in Theorems \ref{thm:consistency}--\ref{thm:gmm_asymptotic_normal}.

Before presenting the GMM estimator, we express the instruments appearing in the moment conditions in matrix form. Let $M^{0}(\boldsymbol\theta)=[\boldsymbol X,\;\lambda\widetilde{\boldsymbol Y}^{*,0}(z),\;\lambda\partial\widetilde{\boldsymbol Y}^{*,0}(z)/\partial z]\in\mathbb R^{n\times(p+2)}$ be the matrix of oracle instruments, and let
 $M(\boldsymbol\theta)=[\boldsymbol X,\;\lambda\widetilde{\boldsymbol Y}^{*}(z),\;\lambda\partial\widetilde{\boldsymbol Y}^{*}(z)/\partial z]\in\mathbb R^{n\times(p+2)}$ be its feasible counterpart. For a given $\boldsymbol\theta$,  define the residual
vector $\boldsymbol e_n(\boldsymbol\theta)=\boldsymbol Y-\lambda\boldsymbol Y^{\dagger}(z)-\boldsymbol X\boldsymbol\beta$, with $e_i(\boldsymbol\theta)$ denoting its $i$-th element.  The corresponding observation-specific feasible moment is given by $\boldsymbol g^{(i)}(\boldsymbol\theta)=[X_i^\top, \lambda\widetilde Y_i^{*}(z),\lambda\partial\widetilde Y_i^{*}(z)/\partial z]^\top e_i(\boldsymbol\theta)\in\mathbb R^{p+2}$.
The feasible and oracle sample moments are defined, respectively, as $\boldsymbol g_n(\boldsymbol\theta)=n^{-1}M(\boldsymbol\theta)^\top\boldsymbol e_n(\boldsymbol\theta)=n^{-1}\sum_{i=1}^n\boldsymbol g^{(i)}(\boldsymbol\theta)$ and $\boldsymbol g_n^{0}(\boldsymbol\theta)=n^{-1}M^{0}(\boldsymbol\theta)^\top\boldsymbol e_n(\boldsymbol\theta)$. Note that the oracle sample moments $\boldsymbol{g}_n^0(\boldsymbol{\theta})$ cannot be employed in practical estimation, though they are critical for deriving the asymptotic theory.

\begin{algorithm}[htbp]
	\caption{Instrument-Smoothed GMM Estimator for the QSAR Model}
	\label{alg:gmm-qsar}
	\begin{algorithmic}[1]
		
		\STATE \textbf{Input:} observed data $(\boldsymbol Y,\boldsymbol X,\boldsymbol A)$, kernel $k$, and bandwidths $\{\tau_i\}_{i=1}^n$.
		
		\STATE \textbf{Output:} instrument-smoothed GMM estimator $\widehat{\boldsymbol\theta}$.
		
		\STATE Compute $\widehat{\boldsymbol\beta}=(\boldsymbol X^\top\boldsymbol X)^{-1}\boldsymbol X^\top\boldsymbol Y$ and set $\widetilde Y_i=X_i^\top\widehat{\boldsymbol\beta}$.
		
		\FOR{$i=1,\cdots,n$}
		\STATE Construct $\widetilde Y_i^{\dagger}(z)$ from
		$\{\widetilde Y_j:a_{ij}=1\}$, and compute
		$\widetilde Y_i^{*}(z)$ and
		$\partial\widetilde Y_i^{*}(z)/\partial z$.
		\ENDFOR
		
		\STATE Define the instrument matrix
		$M(\boldsymbol\theta)
		=[\boldsymbol X,\lambda\widetilde{\boldsymbol Y}^{*}(z),
		\lambda\partial\widetilde{\boldsymbol Y}^{*}(z)/\partial z]$
		and set
		$\boldsymbol g_n(\boldsymbol\theta)
		=n^{-1}M(\boldsymbol\theta)^\top\boldsymbol e_n(\boldsymbol\theta)$.
		
		\STATE Set $\iota\gets1$ and $\widehat{\Xi}_n^{(1)}=\boldsymbol I_{p+2}$.
		
		\WHILE{not converged}
		
		\STATE Obtain $\widehat{\boldsymbol\theta}_{\iota}\gets\arg\min \boldsymbol g_n(\boldsymbol\theta)^\top\{\widehat{\Xi}_n^{(\iota)}\}^{-1}\boldsymbol g_n(\boldsymbol\theta)$.
		
		\STATE Update $\widehat{\Xi}_n^{(\iota+1)}\gets n^{-1}\sum_{i=1}^n\boldsymbol g^{(i)}(\widehat{\boldsymbol\theta}_{\iota})\boldsymbol g^{(i)}(\widehat{\boldsymbol\theta}_{\iota})^\top$.
		
		\STATE Set $\iota\gets\iota+1$.
		
		\ENDWHILE
		
		\STATE \textbf{return} $\widehat{\boldsymbol\theta}=\widehat{\boldsymbol\theta}_{\iota-1}$.
		
	\end{algorithmic}
\end{algorithm}

Using the feasible sample moments $\boldsymbol g_n(\boldsymbol\theta)$, we construct the GMM objective function and define the feasible estimator.
Let $\{\Xi_n\}_{n\geqslant1}$ be a sequence of symmetric positive definite matrices in $\mathbb R^{(p+2)\times(p+2)}$ such that $\Xi_n\xrightarrow{p}\Xi$ for some positive definite matrix $\Xi$, where $\Xi_n^{-1}$ is the GMM weighting matrix. Then the feasible GMM objective function is $J_n(\boldsymbol\theta)=\boldsymbol g_n(\boldsymbol\theta)^\top\Xi_n^{-1}\boldsymbol g_n(\boldsymbol\theta)$, with the corresponding estimator defined as $\widehat{\boldsymbol\theta}=\arg\min_{\boldsymbol\theta\in\boldsymbol\Theta}J_n(\boldsymbol\theta)$.
We next outline a procedure for computing it.

Following the standard iterated GMM procedure \citep{hansen1982large,hayashi2011econometrics}, we first obtain a consistent preliminary estimator $\widehat{\boldsymbol\theta}_1$ using the initial weighting matrix $\{\widehat{\Xi}_n^{(1)}\}^{-1}=\boldsymbol I_{p+2}\in\mathbb R^{(p+2)\times(p+2)}$. For $\iota\geqslant2$ and a given $\widehat{\boldsymbol\theta}_{\iota-1}$, we update the GMM weighting matrix by setting  $\{ \widehat{\Xi}_n^{(\iota)}\}^{-1} = \{ n^{-1}\sum_{i=1}^n\boldsymbol g^{(i)}(\widehat{\boldsymbol\theta}_{\iota-1})\boldsymbol g^{(i)}(\widehat{\boldsymbol\theta}_{\iota-1})^\top\}^{-1}$. Repeating these updates until convergence yields the iterated GMM estimator. The implementation steps are detailed in Algorithm  \ref{alg:gmm-qsar}.


\csection{Theoretical Results}

This section consists of two subsections. Subsection 4.1 establishes the existence and uniqueness of the equilibrium, as well as the local identification of the QSAR model. Subsection 4.2 establishes the consistency of the feasible GMM estimator $\widehat{\boldsymbol\theta}$, introduces the doubly smoothed GMM estimator $\widehat{\boldsymbol\theta}^s$, and derives its consistency and asymptotic normality.

\csubsection{Equilibrium and Local Identification of the QSAR Model}

We begin by showing that the response vector $\boldsymbol Y$ is uniquely determined by the QSAR model.
For any $\boldsymbol y \in \mathbb{R}^n$, let $\boldsymbol y^{\dagger}(z)$ denote the vector with the $i$-th component given by \eqref{eq:quantilenorm}. It suffices to note that, for each fixed $z\in(0,1)$, the mapping $\boldsymbol y\mapsto\boldsymbol y^{\dagger}(z)$ is non-expansive under the infinity norm. That is, for any $\boldsymbol y,\boldsymbol y'\in\mathbb R^n$, $\|\boldsymbol y^{\dagger}(z)-(\boldsymbol y')^{\dagger}(z)\|_\infty\leqslant\|\boldsymbol y-\boldsymbol y'\|_\infty$.
Now, define $Q_0(\boldsymbol y):=\lambda_0\boldsymbol y^{\dagger}(z_0)+\boldsymbol X\boldsymbol\beta_0+\boldsymbol\varepsilon$. It immediately follows that, for any $\boldsymbol y,\boldsymbol y'\in\mathbb R^n$, we have $\|Q_0(\boldsymbol y)-Q_0(\boldsymbol y')\|_\infty\leqslant |\lambda_0|\|\boldsymbol y^{\dagger}(z_0)-(\boldsymbol y')^{\dagger}(z_0)\|_\infty\leqslant |\lambda_0|\|\boldsymbol y-\boldsymbol y'\|_\infty$.
Consequently, for $|\lambda_0|<1$, $Q_0$ is a contraction operator on $(\mathbb R^n,\|\cdot\|_\infty)$. By the contraction mapping theorem \citep{goebel1990topics}, $Q_0$ has a unique fixed point, which establishes the existence and uniqueness of the equilibrium response vector. For   simplicity, we denote the fixed point by $\boldsymbol Y$.

We next establish that, under appropriate distributional and network-propagation conditions, the QSAR responses $\boldsymbol Y$ remain controlled. Unlike LIM models, the QSAR model lacks an explicit linear reduced form because the quantile social norm is determined by the empirical distribution of neighbours' responses.
As a result, node propagation cannot be described by a fixed row-normalized weighting matrix $\boldsymbol W$.
To manage this nonlinear propagation, we define the class of network-related row-normalized matrices as
\[
\mathcal C(\boldsymbol W)
=
\left\{
\boldsymbol B\in\mathbb R^{n\times n}:
B_{ij}\geqslant0,\;
B_{ij}=0\text{ whenever }w_{ij}=0,\;
\sum_{j=1}^nB_{ij}=1
\right\}.
\]
Furthermore, for any $\boldsymbol B,\boldsymbol B_0\in\mathcal C(\boldsymbol W)$, define $H(\boldsymbol B, \boldsymbol B_0)=\boldsymbol B(\boldsymbol I_n-|\lambda_0| \boldsymbol B_0)^{-1}$, $C_{X,n}=\max_{1\leqslant i\leqslant n}\|X_i\|$, and $\kappa_W=\sup_{\boldsymbol B, \boldsymbol B_0\in\mathcal C(\boldsymbol W)}n^{-1}\sum_{\ell=1}^n\{\sum_{i=1}^nH(\boldsymbol B, \boldsymbol B_0)_{i\ell}\}^2$.
The quantity $\kappa _{W}$ measures the cumulative network effect of a shock to a single node's error term
$\varepsilon _{\ell }$ on all other nodes via the quantile social multiplier. For theoretical convenience, we introduce the following assumption.

\noindent \textbf{Assumption 1: (Quantile-Propagation)}
Assume that $\kappa_W=o(n/\log^6 n)$ and $C_{X,n}=O(\log n)$.

\noindent
This assumption is similar to the standard stability restrictions imposed in LIM models, adapted here to the quantile setting. For example, \citet{Lee2004} and \citet{kelejian2010specification} imposed bounded row and column sums for network-related matrices, such as $\|\boldsymbol W\|_1=O(1)$ and $\|(\boldsymbol I_n-\lambda_0 \boldsymbol W)^{-1}\|_1=O(1)$. Because the local dependence structure in our setting varies with the realized ordering of $\boldsymbol Y$, we impose a uniform condition
over $\mathcal C(\boldsymbol W)$. Consequently, the bound
$n^{-1}\sum_{\ell=1}^n\{\sum_{i=1}^n H(\boldsymbol B, \boldsymbol B_0)_{i\ell}\}^2\leqslant \|H( \boldsymbol B, \boldsymbol B_0)\|_1^2\leqslant \|\boldsymbol B\|_1^2\|(\boldsymbol I_n-|\lambda_0| \boldsymbol B_0)^{-1}\|_1^2$
implies that enforcing classical $\ell _{1}$-norm restrictions across these matrices in $\mathcal C(\boldsymbol W)$ yields $\kappa_W=O(1)$, which satisfies our quantile-propagation assumption.
To further control the tail behavior of $Y_i$ under the nonlinear propagation structure, we consider a Log-Sobolev inequality below, which provides the concentration bounds for our subsequent analysis.

\noindent
{\bf{Definition 1}}(Log-Sobolev Inequality \citep{BoucheronLugosiMassart2013})
\label{def:Log_Sobolev}
A probability measure $\mu$ on $\mathbb R^d$  satisfies the Log-Sobolev inequality with constant $c>0$, denoted by $\mathrm{LSI}(c)$, if for every Lipschitz function $f:\mathbb R^d\to\mathbb R$, $\mathrm{Ent}_\mu(f^2)\leqslant 2c\,\E_\mu\|\nabla f\|^2$ holds. Here, the entropy functional, for any non-negative measurable $h$ with $0<\E_\mu h<\infty$, is defined as $\mathrm{Ent}_\mu(h)=\E_\mu\{h\log(h/\E_\mu h)\}$.

\noindent
The LSI characterizes the concentration of $\mu$. For an $L$-Lipschitz function $f$, the LSI implies its moment generating function satisfies $\E_\mu \exp\{s(f-\E_\mu f)\}\leqslant \exp(cL^2s^2/2)$ for any $s\in\mathbb R$. In addition, applying Chernoff's bound yields $\mathrm P_\mu(|f-\E_\mu f|\geqslant t)\leqslant 2\exp\{-t^2/(2cL^2)\}$ for any  $t>0$, ensuring the probability of deviations from the mean decays exponentially.
While the LIM literature assumes that the errors $\{\varepsilon_i\}_{i=1}^n$ have bounded higher-order moments to control concentration \citep{Lee2007,lee2023qml}, the stronger dependence among nodes in the QSAR setting requires a slightly stronger Assumption 2 below, namely that  errors
$\{\varepsilon_i\}_{i=1}^n$  satisfy a Log-Sobolev inequality.

\noindent\textbf{Assumption 2: (Log-Sobolev Noise)}
Assume that $\{\varepsilon_i\}_{i=1}^n$ are independent and zero-mean errors.
Furthermore, suppose the distribution of each $\varepsilon_i$ is absolutely continuous with respect to the Lebesgue measure and satisfies the Log-Sobolev inequality $\mathrm{LSI}(c)$ defined in Definition 1.

Using these two assumptions and the Lipschitz continuity of the quantile function, we obtain Proposition \ref{prop:Quantile_bounded} below.
\begin{proposition}[Upper bound for $\|Y_i-\E Y_i\|_{\psi_2}$]
\label{prop:Quantile_bounded}
Under Assumptions 1--2, each response $Y_i$ satisfies $\|Y_i-\E Y_i\|_{\psi_2}\lesssim (1-|\lambda_0|)^{-1}$.
\end{proposition}
\noindent The detailed proof of Proposition \ref{prop:Quantile_bounded} is provided in Section S2.1 of the supplementary material.
Proposition \ref{prop:Quantile_bounded} shows that the Log-Sobolev inequality provides tail control for each component of the equilibrium response vector.

We next study local identification using the oracle population moment.
We define the oracle population moment as $ \boldsymbol g^{0}(\boldsymbol\theta)=n^{-1}\{M^{0}(\boldsymbol\theta)\}^{\top}\boldsymbol r(\boldsymbol\theta)$, where $\boldsymbol r(\boldsymbol\theta)=\E\{\boldsymbol e_n(\boldsymbol\theta)\}$, with $\boldsymbol e_n(\boldsymbol\theta)$ the residual vector defined in Subsection 3.2.
Let $\boldsymbol\mu(z)=\E\{\boldsymbol Y^{\dagger}(z)\}$. It follows that $\boldsymbol r(\boldsymbol\theta)=\boldsymbol X(\boldsymbol\beta_0-\boldsymbol\beta)+\lambda_0\boldsymbol\mu(z_0)-\lambda\boldsymbol\mu(z)$, which immediately yields $\boldsymbol r(\boldsymbol\theta_0)=\mathbf 0$. For any $z\neq z_0$, we define the secant vector $\boldsymbol v(z,z_0)=\{\boldsymbol\mu(z)-\boldsymbol\mu(z_0)\}/(z-z_0)$ and collect the peer-effect-related terms in matrix $R(z,z_0)=[\boldsymbol\mu(z),\;\lambda_0\boldsymbol v(z,z_0)]\in\mR^{n\times 2}$.
Let $Z_P^0(\lambda,z)=[\lambda\widetilde{\boldsymbol Y}^{*,0}(z),\;\lambda\partial\widetilde{\boldsymbol Y}^{*,0}(z)/\partial z]$, and $P_X^\perp={\boldsymbol I_n}-\boldsymbol X(\boldsymbol X^\top\boldsymbol X)^{-1}\boldsymbol X^\top$. To assess instrument relevance after removing the covariate contribution, we introduce the partial secant relevance matrix, $\mathcal S_{\tau}^{0,\mathrm{sec}}(\lambda,z)=n^{-1}\{Z_P^0(\lambda,z)\}^\top P_X^\perp R(z,z_0)$, and make the following assumption. Here,
$\lambda_{\min}(\cdot)$ and $\sigma_{\min}(\cdot)$ denote the minimum eigenvalue and the minimum singular value, respectively.

\noindent\textbf{Assumption 3: (Local Secant Relevance)}
There exists a compact neighborhood $\mathcal N\subset\boldsymbol\Theta$ of
$\boldsymbol\theta_0$ and finite positive constants $c_X$, $c_P$ such that,
for all sufficiently large $n$,
$\lambda_{\min}(n^{-1}\boldsymbol X^\top\boldsymbol X)\geqslant c_X$
and
$\sigma_{\min}\{\mathcal S_{\tau}^{0,\mathrm{sec}}(\lambda,z)\}
\geqslant c_P$
uniformly over
$\boldsymbol\theta \in\mathcal N$ with $z\neq z_0$.

\noindent
The first part of Assumption 3 guarantees the identification of $\boldsymbol\beta$, whereas the second ensures that the oracle peer-effect instruments distinguish local changes in $\lambda$ and $z$ after partially removing the influence of $\boldsymbol X$. These conditions align with the standard full-rank requirement for the population moment Jacobian, which is widely adopted in the literature \citep{hansen1982large,chen2014local}.  However, since the quantile  social norm is only piecewise smooth in $z$, $\E\{\boldsymbol Y^{\dagger}(z)\}$ may not be differentiable at $z_0$. This motivates us to formulate the relevance condition using the secant vector $\boldsymbol v(z,z_0)$ of $\boldsymbol\mu(\cdot)$ between $z_0$ and $z$.
The subsequent proposition establishes oracle local identification of the true parameter $\boldsymbol\theta_0$. The proof is provided in Section S2.2 of the supplementary material.

\begin{proposition}[Oracle Local Identification]
\label{prop:identification}
Suppose Assumptions 1--3 hold. Then, for all sufficiently large $n$,
$\boldsymbol g^{0}(\boldsymbol\theta)=\mathbf0$ for
$\boldsymbol\theta\in\mathcal N$ if and only if
$\boldsymbol\theta=\boldsymbol\theta_0$, where $\mathcal N$ is the compact neighborhood specified in Assumption 3.
\end{proposition}

\csubsection{Asymptotic Theory of the QSAR Model}

In this subsection, we develop the asymptotic theory of the QSAR estimator. We first establish the consistency of the feasible estimator $\widehat{\boldsymbol\theta}$. To derive asymptotic normality, we smooth the quantile social norm of the residual and construct the smoothed moment objective function $J_n^s(\boldsymbol\theta)$. For each node $i$, let $\widetilde Y_{i,(1)}^0\leqslant\cdots\leqslant\widetilde Y_{i,(d_i)}^0$ denote the ordered oracle pseudo responses of its neighbours, and define $\widetilde\Delta_{ik}^0=\widetilde Y_{i,(k+1)}^0-\widetilde Y_{i,(k)}^0$. To formally derive these asymptotic properties, we impose the following set of regularity conditions.

\noindent\textbf{Assumption 4: (Kernel Regularity)}
The kernel function $k(\cdot)$ is nonnegative, symmetric, and twice continuously differentiable. It satisfies $\int k(u)\,du=1$, $\int uk(u)\,du=0$, $\int u^2k(u)\,du > 0$, and, for some $\nu>0$, $\int |u|^{2+\nu}k(u)\,du<\infty$. Moreover, $\sup_u\big\{|k(u)|+|k'(u)|+|k''(u)|\big\}<\infty$ and
$\int(1+|u|^{2+\nu})\big\{|k'(u)|+|k''(u)|\big\}\,du<\infty$.

\noindent\textbf{Assumption 5: (Local Spacing and Stability)}
\textnormal{(i)} There exists a finite positive constant $C_\Delta$ such that $\sup_{i,k} d_i\widetilde\Delta_{ik}^0\leqslant C_\Delta$.
\textnormal{(ii)} There exists a finite positive constant $C_M<\infty$ such that $\lambda_{\max}(Q_X)\leqslant C_M$, $n^{-1}\|Z_P^0(\lambda,z)\|_{\mathrm F}^2\leqslant C_M$, and $n^{-1}\|R(z,z_0)\|_{\mathrm F}^2\leqslant C_M$ uniformly over $\boldsymbol\theta\in\mathcal N$ with $z\neq z_0$, where $\mathcal N$ is the compact neighborhood specified in Assumption 3.

\noindent\textbf{Assumption 6: (Instrument Bandwidth Regularity)}
The instrument-smoothing bandwidths satisfy
$\tau_i\gg\max\{(\log n)^2(1+\kappa_W)/\sqrt n,\;(\log n)^{3/2}(1+\kappa_W)^{1/4}/n^{1/4}\}$ and $\tau_i \to 0$
uniformly over $1\leqslant i\leqslant n$.

\noindent
Assumption 4 imposes a standard second-order smooth kernel condition with a weak tail-moment restriction, which is readily satisfied by the Gaussian kernel.
Assumption 5(i) is a mild regularity condition on the covariates, requiring the oracle pseudo responses $\widetilde Y_j^0$ of each node's neighbours $j\in\mathcal A_i$ to exhibit uniformly bounded local gaps $\widetilde\Delta_{ik}^0$. Assumption 5(ii) imposes uniform boundedness on the covariates, instruments, and residual secants.  Assumption 6 introduces a lower bound on instrument smoothing bandwidths so that the feasible-oracle discrepancy is asymptotically negligible. Under the aforementioned assumptions, Theorem \ref{thm:consistency} establishes consistency of the GMM estimator.

\begin{theorem}[Consistency of $\widehat{\boldsymbol\theta}$]
\label{thm:consistency}
Suppose Assumptions 1--6 hold. Let $\widehat{\boldsymbol\theta}$ be a minimizer of $J_n(\boldsymbol\theta)$ over the neighbourhood $\mathcal N$ in Proposition \ref{prop:identification}. Then, as $n\to\infty$, we have $\widehat{\boldsymbol\theta}\xrightarrow{p}\boldsymbol\theta_0$.
\end{theorem}

The proof of Theorem \ref{thm:consistency} is provided in Section S2.3 of the supplementary material.
Next, we establish asymptotic normality. The consistency result above relies on uniform convergence and local identification and thus does not require the objective function to be differentiable. In contrast, the standard GMM argument for asymptotic normality uses the first-order condition together with a Taylor expansion of the sample moment around $\boldsymbol\theta_0$. Although instrument smoothing is used in $M(\boldsymbol\theta)$ to construct smooth and feasible instruments, the sample moment $\boldsymbol g_n(\boldsymbol\theta)$ is not continuously differentiable because its residual contains the piecewise-linear quantile social norm $\boldsymbol Y^{\dagger}(z)$; in particular, the derivative of each $Y_i^{\dagger}(z)$ varies over interpolation knots and is undefined at the knots themselves. Thus, smoothing the instruments alone cannot resolve the sample moment's nonsmoothness, making the standard GMM argument still inapplicable.

To address this issue, we further smooth $Y_i^{\dagger}(z)$ in the residual.
Specifically, for each node $i$, we replace $Y_i^{\dagger}(z)$ in the residual with its smoothed version $Y_i^{*}(z)=\int k(u)Y_{i,\mathrm{ext}}^{\dagger}(z-h_i u)\,du$, where $Y_{i,\mathrm{ext}}^{\dagger}(v)$ denotes the constant extension of $Y_i^{\dagger}$, taking the values $\min_{j\in\mathcal A_i}Y_j$ for $v\leqslant0$ and $\max_{j\in\mathcal A_i}Y_j$ for $v\geqslant1$. The node-specific residual-smoothing bandwidth $h_i$ is distinct from the instrument-smoothing bandwidth $\tau_i$. This residual smoothing makes the resulting sample moment differentiable in $z$, allowing the first-order condition and Taylor expansion to be applied. Stacking these terms into $\boldsymbol Y^{*}(z)$, we substitute the raw residual $\boldsymbol e_n(\boldsymbol\theta)=\boldsymbol Y-\lambda\boldsymbol Y^{\dagger}(z)-\boldsymbol X\boldsymbol\beta$ with its smoothed counterpart $\boldsymbol e_n^s(\boldsymbol\theta)=\boldsymbol Y-\lambda\boldsymbol Y^{*}(z)-\boldsymbol X\boldsymbol\beta$, and subsequently define the so-called \textit{doubly smoothed GMM estimator} $\widehat{\boldsymbol\theta}^s$ below.

We first lay out necessary notation prior to formally introducing $\widehat{\boldsymbol\theta}^s$.
Recalling the ordered neighbor responses $Y_{i,(1)},\cdots,Y_{i,(d_i)}$ defined in Subsection 2.2, let $m_{ik}=\E\{Y_{i,(k)}\}$ for $1\leqslant k\leqslant d_i$ and set $s_{ik}=(d_i-1)(m_{i,k+1}-m_{ik})$ for $1\leqslant k\leqslant d_i-1$.
Since $\mathcal Z=[\delta,1-\delta]$, define
$\mathcal K_i^s=\{k\in\{1,\cdots,d_i-1\}:k/(d_i-1)\geqslant\delta,\ (k-1)/(d_i-1)\leqslant1-\delta\}$
and
$\mathcal K_i^\Delta=\{k\in\{1,\cdots,d_i-2\}:\delta\leqslant k/(d_i-1)\leqslant1-\delta\}$, which identify the interpolation intervals $[t_{ik},t_{i,k+1}]$ intersecting $\mathcal Z$ and the interpolation knots $t_{i,k+1}$ within $\mathcal Z$, respectively, and are used in Assumption 7(i) to bound the slopes $s_{ik}$ and their changes $s_{i,k+1}-s_{ik}$.
Let $e_i^s(\boldsymbol\theta)$ denote the $i$-th element of $\boldsymbol e_n^s(\boldsymbol\theta)$ and define the observation-specific feasible smoothed moment as $\boldsymbol g^{s,(i)}(\boldsymbol\theta)=[X_i^\top,\lambda\widetilde Y_i^{*}(z),\lambda\partial\widetilde Y_i^{*}(z)/\partial z]^\top \allowbreak e_i^s(\boldsymbol\theta)$. The feasible and oracle smoothed sample moments are $\boldsymbol g_n^s(\boldsymbol\theta)=n^{-1}M(\boldsymbol\theta)^\top\boldsymbol e_n^s(\boldsymbol\theta)=n^{-1}\sum_{i=1}^n\boldsymbol g^{s,(i)}(\boldsymbol\theta)$ and $\boldsymbol g_n^{s,0}(\boldsymbol\theta)=n^{-1}\{M^0(\boldsymbol\theta)\}^\top\boldsymbol e_n^s(\boldsymbol\theta)$, respectively.
We then set $J_n^s(\boldsymbol\theta)=\boldsymbol g_n^s(\boldsymbol\theta)^\top \allowbreak \Xi_n^{-1}\boldsymbol g_n^s(\boldsymbol\theta)$, and define the \textit{doubly smoothed GMM estimator} $\widehat{\boldsymbol\theta}^s$ to be a minimizer of $J_n^s(\boldsymbol\theta)$ over the compact neighbourhood $\mathcal N$ introduced in Proposition \ref{prop:identification}.

Substituting $\boldsymbol{e}_n^s(\boldsymbol{\theta})$ in place of the original residual $\boldsymbol{e}_n(\boldsymbol{\theta})$ induces smoothing bias across all three oracle population moment conditions. To capture this bias, we introduce the pseudo-parameter $\boldsymbol\theta_h^{s,0}$.
Define the population smoothed oracle moment
$\boldsymbol{g}_h^{s,0}(\boldsymbol{\theta})=n^{-1}\{M^0(\boldsymbol{\theta})\}^\top\mathbb{E}\big\{\boldsymbol{e}_n^s(\boldsymbol{\theta})\big\}$
and its corresponding population objective
$J_h^{s,0}(\boldsymbol{\theta})=\boldsymbol{g}_h^{s,0}(\boldsymbol{\theta})^\top\Xi^{-1}\boldsymbol{g}_h^{s,0}(\boldsymbol{\theta}),$
and let $\boldsymbol{\theta}_h^{s,0}$ denote a minimizer of $J_h^{s,0}(\boldsymbol{\theta})$ over $\mathcal{N}$.
Since $\mathcal{N}$ is compact and $J_h^{s,0}(\boldsymbol{\theta})$ is continuous, such a minimizer is guaranteed to exist. Even though the smoothed population moment is generally nonzero at $\boldsymbol\theta_0$, the bandwidth restrictions stated in Assumption 7 below ensure that the resulting smoothing bias is negligible at the $n^{-1/2}$ order.

\noindent\textbf{Assumption 7: (Response Regularity and Residual-smoothing Bandwidth)}
\textnormal{(i)} There exists a positive constant $C_Y<\infty$ such that $|s_{ik}|\leqslant C_Y$ uniformly over $1\leqslant i\leqslant n$ and $k\in\mathcal K_i^s$, and $d_i|s_{i,k+1}-s_{ik}|\leqslant C_Y$ uniformly over $1\leqslant i\leqslant n$ and $k\in\mathcal K_i^\Delta$. \textnormal{(ii)} The node-specific residual-smoothing bandwidths $\{h_i\}_{i=1}^n$ satisfy $\max\{(\log n)^{3/2}\sqrt{(1+\kappa_W)/n}, \allowbreak d_i^{-1}\}\ll h_i\ll n^{-1/4}$ uniformly over $1\leqslant i\leqslant n$.

\noindent
Intuitively, Assumption 7(i) requires the expected quantile social norm $\E\{Y_i^{\dagger}(z)\}$ to change smoothly across the interpolation knots with uniformly controlled local gaps. Section S5 of the supplementary material shows that Assumption 7(i) is satisfied in several representative examples. Assumption 7(ii) specifies lower and upper bounds on the residual-smoothing bandwidths $h_i$ to control estimation variance and smoothing bias, respectively. The following proposition formalizes this result by showing that the difference between the pseudo parameter $\boldsymbol\theta_{h}^{s,0}$ and $\boldsymbol\theta_0$ is negligible at the $\sqrt n$ scale.

\begin{proposition}
\label{prop:bias}
Under Assumptions 1--7, as $n\to\infty$,
$\|\boldsymbol\theta_h^{s,0}-\boldsymbol\theta_0\|
=o(n^{-1/2})$.
\end{proposition}

\noindent
The proof of Proposition \ref{prop:bias} is provided in Section S2.4 of the supplementary material. This result shows that the population bias induced by kernel smoothing is negligible at the $\sqrt n$ scale. We next establish the consistency of the doubly smoothed GMM estimator. The proof of Theorem \ref{thm:smoothed_gmm_consistency} is provided in Section S2.5 of the supplementary material.

\begin{theorem}[Consistency of $\widehat{\boldsymbol\theta}^s$]
\label{thm:smoothed_gmm_consistency}
Suppose Assumptions 1--7 hold, and let $\widehat{\boldsymbol\theta}^s$ minimize $J_n^s$ over $\mathcal N$ in Proposition \ref{prop:identification}. Then, as $n\to\infty$, $\widehat{\boldsymbol\theta}^s\xrightarrow{p}\boldsymbol\theta_0$.
\end{theorem}

We next introduce the notation required to establish asymptotic normality.
Define $D_n^{s,0}(\boldsymbol\theta)=\partial\boldsymbol g_n^{s,0}(\boldsymbol\theta)/\partial\boldsymbol\theta^\top$ and $G_h^{s,0}=\E\{D_n^{s,0}(\boldsymbol\theta_0)\}$. Let $\operatorname{dist}_n(i,j)$ denote the shortest-path distance between nodes $i$ and $j$ in the undirected graph obtained by ignoring edge directions. For nonnegative integers $r$ and $m$, define $\mathcal N_n(i;r)=\{j:\operatorname{dist}_n(i,j)\leqslant r\}$ and $\mathcal N_n^\partial(i;m)=\{j:\operatorname{dist}_n(i,j)=m\}$. For $v>0$, define $\delta_n^\partial(m;v)=n^{-1}\sum_{i=1}^n|\mathcal N_n^\partial(i;m)|^v$ and $\Delta_n^{\mathrm{net}}(m,r;v)=n^{-1}\sum_{i=1}^n\max_{j\in\mathcal N_n^\partial(i;m)}|\mathcal N_n(i;r)\setminus\mathcal N_n(j;m-1)|^v$, where $\mathcal N_n(j;-1)=\varnothing$ and the inner maximum is set to zero when $\mathcal N_n^\partial(i;m)=\varnothing$. For $\ell>0$, define $c_n(m,r;\ell)=\inf_{\alpha>1}\{\Delta_n^{\mathrm{net}}(m,r;\ell\alpha)\}^{1/\alpha}\allowbreak\,\{\delta_n^\partial(m;\alpha')\}^{1/\alpha'}$, where $\alpha'=\alpha/(\alpha-1)$. The neighbourhood growth measure $c_n(m,r;\ell)$ captures how neighbourhood sizes grow with network distance by considering the average number of nodes within distance $r$ and at distance $m$. Let $h_{\max,n}=\max_{1\leqslant i\leqslant n}h_i$, $\bar\lambda=1-c_\lambda$, $\vartheta_{n,0}=1$, and $\vartheta_{n,m}=h_{\max,n}\wedge\bar\lambda^{\lfloor(m-1)/2\rfloor}$ for $m\geqslant1$. The dependency upper bound $\vartheta_{n,m}$ characterizes the dependence between two nodes at distance $m$, while the sequence $\{\vartheta_{n,m}\}_{m\geqslant0}$ captures its decay as the distance increases. The following assumptions impose the moment and network conditions required for asymptotic normality.

\noindent\textbf{Assumption 8: (Oracle Moment Regularity)}
\textnormal{(i)} The covariance matrix of the smoothed oracle moment function satisfies
$\operatorname{Var}\{\sqrt n\,\boldsymbol g_n^{s,0}(\boldsymbol\theta_0)\}
=\boldsymbol\Sigma_n\to\boldsymbol\Sigma$
with $\lambda_{\min}(\boldsymbol\Sigma)>0$.
\textnormal{(ii)} The expected Jacobian satisfies
$G_h^{s,0}\to G_0$ with $\operatorname{rank}(G_0)=p+2$.

\noindent\textbf{Assumption 9: (Neighbourhood Growth and Dependence Decay)}
For some $q>4$ and some integer sequence $r_n\to\infty$, we have $n^{3/2}(\log n)\vartheta_{n,r_n}^{1-1/q}\to0$ and $(\log n)^{2+\ell}n^{-\ell/2}\allowbreak \sum_{m\geqslant0}c_n(m,r_n;\ell)\vartheta_{n,m}^{1-(2+\ell)/q}\to0$ for $\ell=1,2$.

\noindent
Assumption 8 ensures that the asymptotic covariance matrix $\boldsymbol\Sigma$ of $\sqrt n\,\boldsymbol g_n^{s,0}(\boldsymbol\theta_0)$ is nondegenerate and the limiting Jacobian $G_0$ has full rank. Such conditions are widely adopted in GMM asymptotic theory for spatial and network models \citep{hansen1982large,lee2010efficient,kuersteiner2020dynamic}. Assumption 9 requires the dependency upper bound $\vartheta_{n,r_n}$ at distance $r_n$ to decay sufficiently fast, and balances neighbourhood growth $c_n(m,r_n;\ell)$ against dependence decay $\vartheta_{n,m}$ across all distances. Assumption 9 aligns with Assumption 3.4 of \citet{kojevnikov2021limit}. We present the asymptotic normality result for the doubly smoothed GMM estimator in Theorem \ref{thm:gmm_asymptotic_normal}.

\begin{theorem}[Asymptotic normality]
\label{thm:gmm_asymptotic_normal}
Suppose Assumptions 1--9 hold. Then, as $n\to\infty$, $\sqrt n(\widehat{\boldsymbol\theta}^s-\boldsymbol\theta_0)\xrightarrow{d}N(\mathbf0_{(p+2)\times1},\boldsymbol V)$, where $\boldsymbol V=(G_0^\top\Xi^{-1}G_0)^{-1}G_0^\top\Xi^{-1}\boldsymbol\Sigma\Xi^{-1}G_0(G_0^\top\Xi^{-1}G_0)^{-1}$.
\end{theorem}

\noindent
The proof is provided in Supplementary Section S2.6. We first establish the asymptotic normality of the network-dependent oracle moment function $\sqrt n\,\boldsymbol g_n^{s,0}(\boldsymbol\theta_0)$ and then show that the feasible moment $\boldsymbol g_n^s(\boldsymbol\theta_0)$ is asymptotically equivalent to its oracle counterpart.
We use this result to demonstrate the asymptotic normality of the doubly smoothed GMM estimator $\widehat{\boldsymbol\theta}^s$. Its updated estimation procedure is presented in Algorithm \ref{alg:gmm-qsar-smoothed}. Relative to Algorithm \ref{alg:gmm-qsar}, the additional step is to smooth the quantile social norm $Y_i^{\dagger}(z)$ using $h_i$.

\begin{algorithm}[htbp]
\caption{Doubly Smoothed GMM Estimator for the QSAR Model}
\label{alg:gmm-qsar-smoothed}
\begin{algorithmic}[1]

\STATE \textbf{Input:} observed data $(\boldsymbol Y,\boldsymbol X,\boldsymbol A)$, kernel $k$, and bandwidths $\{h_i,\tau_i\}_{i=1}^n$.

\STATE \textbf{Output:} doubly smoothed GMM estimator $\widehat{\boldsymbol\theta}^s$.

\STATE Compute $\widehat{\boldsymbol\beta}=(\boldsymbol X^\top\boldsymbol X)^{-1}\boldsymbol X^\top\boldsymbol Y$ and set $\widetilde Y_i=X_i^\top\widehat{\boldsymbol\beta}$.

\FOR{$i=1,\cdots,n$}
\STATE Construct $Y_i^{\dagger}(z)$ from $\{Y_j:a_{ij}=1\}$ and compute $Y_i^{*}(z)$ with bandwidth $h_i$.
\STATE Construct $\widetilde Y_i^{\dagger}(z)$ from $\{\widetilde Y_j:a_{ij}=1\}$, and compute $\widetilde Y_i^{*}(z)$ and $\partial\widetilde Y_i^{*}(z)/\partial z$ with bandwidth $\tau_i$.
\ENDFOR

\STATE Define the smoothed residual vector
$\boldsymbol e_n^s(\boldsymbol\theta)
=\boldsymbol Y-\lambda\boldsymbol Y^{*}(z)-\boldsymbol X\boldsymbol\beta$, the instrument matrix
$M(\boldsymbol\theta)
=[\boldsymbol X,\lambda\widetilde{\boldsymbol Y}^{*}(z),
\lambda\partial\widetilde{\boldsymbol Y}^{*}(z)/\partial z]$
and set
$\boldsymbol g_n^s(\boldsymbol\theta)
=n^{-1}M(\boldsymbol\theta)^\top\boldsymbol e_n^s(\boldsymbol\theta)$.

\STATE Set $\iota\gets1$ and $\widehat{\Xi}_n^{(1)}=\boldsymbol I_{p+2}$.

\WHILE{not converged}

\STATE Obtain $\widehat{\boldsymbol\theta}_{\iota}^s\gets\arg\min \boldsymbol g_n^s(\boldsymbol\theta)^\top\{\widehat{\Xi}_n^{(\iota)}\}^{-1}\boldsymbol g_n^s(\boldsymbol\theta)$.

\STATE Update $\widehat{\Xi}_n^{(\iota+1)}\gets n^{-1}\sum_{i=1}^n\boldsymbol g^{s,(i)}(\widehat{\boldsymbol\theta}_{\iota}^s)\boldsymbol g^{s,(i)}(\widehat{\boldsymbol\theta}_{\iota}^s)^\top$.

\STATE Set $\iota\gets\iota+1$.

\ENDWHILE

\STATE \textbf{return} $\widehat{\boldsymbol\theta}^s=\widehat{\boldsymbol\theta}_{\iota-1}^s$.

\end{algorithmic}
\end{algorithm}

\noindent\textbf{Remark: (Testing for the Presence of Peer Effects)}
Identification of the quantile level $z_0$ requires $\lambda_0\neq0$. If $\lambda_0=0$, the quantile social norm term vanishes and $z_0$ becomes an unidentified nuisance parameter. Therefore, before interpreting the estimated quantile social norm, it is necessary to test $\mathrm H_0:\lambda_0=0$ against $\mathrm H_1:\lambda_0\neq0$. Since standard Wald-type tests are not directly applicable when $z_0$ is unidentified under the null \citep{Davies1987}, Section S4 of the supplementary material develops a grid-based testing procedure inspired by \citet{AndrewsPloberger1994} and \citet{Hansen1996}. Specifically, for each candidate quantile level $z$, we first formulate fixed-$z$ oracle instruments using pseudo responses that are independent of the errors $\boldsymbol\varepsilon$. The orthogonality yields moment conditions for identifying $(\boldsymbol\beta^\top,\lambda)^\top$ conditional on this fixed $z$.
The asymptotic normality of the resulting fixed-$z$ estimator under $\mathrm H_0$ is carefully established.
Building on this fixed-$z$ asymptotic theory, we construct a statistic which aggregates information across a pre-specified grid of quantile levels and develop a corresponding residual bootstrap procedure.

\csection{Numerical Analysis}

\csubsection{Simulation Studies}

We evaluate the proposed model under three network structures and network sizes \(n\in\{1000,2000,3000\}\). Each configuration is replicated \(R=500\) times. For a given adjacency matrix \(\boldsymbol A\), the response \(\boldsymbol Y\) is the unique fixed point \(\boldsymbol Y=\lambda_0\boldsymbol Y^{\dagger}(z_0)+\boldsymbol X\boldsymbol\beta_0+\boldsymbol\varepsilon\), and we iterate this mapping until \(\|\boldsymbol Y^{(t+1)}-\boldsymbol Y^{(t)}\|_\infty<10^{-9}\). We set \(\boldsymbol\beta_0=(0.90,-0.65)^\top\), \(\lambda_0=0.28\), \(z_0\in\{0.3,0.5,0.7\}\), and independently draw \(\varepsilon_i\sim N(0,0.3^2)\).
For each of the three network designs, we independently draw $U_i,V_i\sim N(0,1)$, define $U_i^{\mathrm{net}}=\sum_j a_{ij}U_j/\sqrt{d_i}$ and $V_i^{\mathrm{net}}=\sum_j a_{ij}(V_j^2-1)/\sqrt{2d_i}$, and construct the covariates as $X_{i1}=(U_i+\rho U_i^{\mathrm{net}})/\sqrt{1+\rho^2}$ and $X_{i2}=(V_i+\rho V_i^{\mathrm{net}})/\sqrt{1+\rho^2}$, where $\rho=1.5$. This construction introduces transparent network dependence while retaining zero means and variances equal to one. We use the standard normal density $k(u)=(2\pi)^{-1/2}\exp(-u^2/2)$ as the kernel function. The residual-smoothing bandwidth is \(h_i=0.06(n/1000)^{-0.32} (d_i/\bar d_n)^{-0.08}\), while the instrument-smoothing bandwidth is \(\tau_i=0.16(n/1000)^{-0.18}(d_i/\bar d_n)^{-0.08}\), where $\bar d_n=n^{-1}\sum_{i=1}^n d_i$.

\textbf{Example 1} (Dyad Independence Network). Following \citet{Holland01031981}, we generate the dyads independently. For each \(i<j\), we set \((a_{ij},a_{ji})=(1,1)\) with probability \(p_n= d^{\mathrm{E}}_n/(n-1)\), where \( d^{\mathrm{E}}_n=64(n/1000)^{0.38}\), and otherwise set \((a_{ij},a_{ji})=(0,0)\). Thus, the network is sparse and has a diverging expected degree.

\textbf{Example 2} (Stochastic Block Network). Following \cite{wang1987stochastic} and \cite{nowicki2001estimation}, we generate a stochastic block model (SBM) network by dividing the nodes into two equally sized blocks. Conditional on block membership, edges are independent: two nodes in the same block are connected with probability $p_{\mathrm{in}}$, while nodes in different blocks are connected with probability $p_{\mathrm{out}}$. We set $p_{\mathrm{in}}=4p_{\mathrm{out}}$ and $p_{\mathrm{out}}=d_n^{\mathrm E}/\{4(n/2-1)+n/2\}$, where $d_n^{\mathrm E}=64(n/1000)^{0.38}$.

\textbf{Example 3} (Power-Law Network). Motivated by power-law degree distributions \citep{Barabsi1999EmergenceOS,Clauset2009}, we set \(d_{n,\min}=\lceil32(n/1000)^{0.38}\rceil\) and \(d_{n,\max}=\lfloor128(n/1000)^{0.44}\rfloor\). We first draw the target degrees \(d_1,\cdots,d_n\) independently from the truncated discrete power-law distribution \(\Pr(d_i=u)=c_{\text{PL}} u^{-2.20}\) for \(u \in\{d_{n,\min},\cdots,d_{n,\max}\}\), where \(c_{\text{PL}}=\{\sum_{\ell=d_{n,\min}}^{d_{n,\max}}\ell^{-2.20}\}^{-1}\) is the normalizing constant. We then use the Viger--Latapy algorithm to generate the undirected graph with the resulting degree sequence.

For \(j=1,\cdots,p+2\), denote the true \(j\)-th parameter by \(\theta_{0j}\), its doubly smoothed GMM estimate in replication \(r\) by \(\widehat\theta_{rj}\), and the corresponding Monte Carlo mean by \(\bar\theta_j=R^{-1}\sum_{r=1}^R\widehat\theta_{rj}\). We report the root mean squared error (RMSE), absolute bias (Bias), and Monte Carlo standard deviation (SE), defined as \(\mathrm{RMSE}_j=\{R^{-1}\sum_{r=1}^R(\widehat\theta_{rj}-\theta_{0j})^2\}^{1/2}\), \(\mathrm{Bias}_j=|R^{-1}\sum_{r=1}^R(\widehat\theta_{rj}-\theta_{0j})|\), and \(\mathrm{SE}_j=\{R^{-1}\sum_{r=1}^R(\widehat\theta_{rj}-\bar\theta_j)^2\}^{1/2}\).
For each replication \(r\), let \(\widehat{\boldsymbol V}_r\) denote the plug-in estimator of the asymptotic covariance matrix \(\boldsymbol V\) in Theorem~\ref{thm:gmm_asymptotic_normal}, obtained by replacing the population quantities in \(\boldsymbol V\) with their sample counterparts. The asymptotic standard error of the \(j\)-th parameter is $\widehat s_{rj}=\{[\widehat{\boldsymbol V}_r]_{jj}/n\}^{1/2}$, and its Monte Carlo average is reported as the theoretical standard error (tSE) given by $\mathrm{tSE}_j=R^{-1}\sum_{r=1}^R\widehat s_{rj}$. The corresponding \(95\%\) interval is \(\mathcal I_{rj}=[\widehat\theta_{rj}-\Phi^{-1}(0.975)\widehat s_{rj},\,\widehat\theta_{rj}+\Phi^{-1}(0.975)\widehat s_{rj}]\), where $\Phi$ denotes the standard normal distribution function, and its empirical coverage probability is \(\mathrm{CP}_j=R^{-1}\sum_{r=1}^R\boldsymbol 1\{\theta_{0j}\in\mathcal I_{rj}\}\).

Table~\ref{tab:network_metrics_0.5_new} reports the results for \(z_0=0.5\); the results for \(z_0=0.3\) and \(z_0=0.7\) are provided in Section~S3 of the supplementary material. The estimates exhibit only small biases, and both RMSE and SE decrease with \(n\) for every parameter and network design. The empirical SE and tSE are close, and the CP values are centred near \(0.95\), ranging from \(0.934\) to \(0.976\). Overall, these results provide finite-sample evidence for the consistency of the estimator and the asymptotic normality established in Theorem~\ref{thm:gmm_asymptotic_normal}.

\csubsection{Real Data Analysis}

This analysis utilizes transaction data from fast-food restaurants in Guangzhou, China, provided by Shouqianba, a leading aggregate-payment platform.\footnote{http://www.shouqianbao.com.cn/} The dataset consists of restaurant-level aggregated information. Specifically, transactions from January 1 to July 1, 2024, are used to construct the response variable, while data from July 1 to December 31, 2023, are employed to construct transaction-related covariates. All variables were pre-processed by the platform and provided as aggregated measurements at the restaurant level. The adjacency matrix is constructed based on geographic distance: for any two restaurants $i$ and $j$, $a_{ij}=a_{ji}=1$ if they are located within 200 metres of each other, and $a_{ij}=a_{ji}=0$ otherwise. The original dataset contains 1,997 restaurants; after removing 456 isolated nodes, 1,541 restaurants and 5,205 undirected edges remain, and the 1,229 restaurants with $d_i\geqslant 2$ are included in the estimation. The primary objective is to evaluate the relationship between peer interactions and restaurant performance.

For each restaurant $i$ ($1\leqslant i\leqslant n$), the response variable $Y_i$ is the processed average daily transaction volume during the observation period and is used directly in the estimation. We consider five covariates: (a) \textit{repeat customer ratio} ($X_{i1}$), defined as the percentage of customers with two or more visits in the previous period; (b) \textit{commercial zone diversity} ($X_{i2}$), measured as the number of distinct commercial zone types surrounding the restaurant (e.g., office, residential, hotel, school, hospital, dining, shopping, and tourist areas); (c) \textit{menu variety} ($X_{i3}$), measured by the number of distinct types of dishes offered; (d) \textit{average transaction value} ($X_{i4}$), the mean transaction amount in the previous period, serving as a proxy for the restaurant's price level; and (e) \textit{operating hours} ($X_{i5}$), a binary variable where $X_{i5}=1$ if the $i$-th restaurant operates during both lunch and dinner hours, and $X_{i5}=0$ otherwise. The repeat customer ratio, commercial zone diversity, and menu variety are standardized to have zero mean and unit variance. Average transaction value is first logarithmically transformed and then standardized, while operating hours is retained as a binary variable.

Prior to conducting the QSAR estimation, we perform a test for the existence of peer effects as described in the supplementary material. Specifically, we employ $B=499$ residual-bootstrap replications over a grid of quantile levels $\{0.1,0.2,\cdots,0.9\}$. The bootstrap $p$-value is calculated as $\widehat p_B=\{1+\sum_{b=1}^{B}\boldsymbol 1(T_K^{(b)}\geqslant T_K)\}/(B+1)$, where $T_K^{(b)}$ denotes the test statistic obtained from the $b$-th bootstrap sample. The observed statistic $T_K=480.672$ substantially exceeds the bootstrap 95\% critical value of 34.747. This result provides strong empirical evidence against $\mathrm  H_0:\lambda_0=0$.

For comparison, we consider three models: the LIM model \citep{Manski1993}, the CES model \citep{EconometricaPeerEffects}, and the QSAR model proposed in this work. As the CES model requires positive responses, we shift the response by a constant of 3.035. Table~\ref{tb:res} reports the estimation results.
The main results are consistent across the three models. Although CES and QSAR use different forms of social norms, the covariate effects are generally stable, and the negative and significant estimates of $\lambda$ provide strong evidence of competitive peer effects among neighbouring restaurants.

To further explore heterogeneity in peer interactions, we assign each restaurant to either the \textit{Low} or \textit{High} group according to its pseudo response $\widetilde Y_i$, using the median of $\{\widetilde Y_i\}_{i=1}^n$ as the cutoff. Because the feasible pseudo response $\widetilde Y_i$ approximates the exogenous oracle pseudo response $\widetilde Y_i^0$, we adopt it rather than the endogenous observed response $Y_i$ for group assignment.
For each group, we follow the procedure described above to estimate the QSAR model. Table~\ref{tab:qsar_two_groups} reports the results. The network coefficients remain negative and significant in both groups. The estimated quantile level is higher in the High group than in the Low group. This result suggests that restaurants with higher transaction volumes are influenced by neighbours at a higher quantile of the transaction-volume distribution.
The covariate results show both common patterns and group differences. The repeat customer ratio, commercial zone diversity, and average transaction value are positive and significant in both groups, indicating that customer retention, the surrounding commercial environment, and price level are closely related to restaurant performance across the two groups. Menu variety is significant only in the Low group, suggesting that a broader menu is more relevant to the performance of restaurants with lower transaction volumes. In contrast, operating hours is significant only in the High group, suggesting that extending operating hours is more relevant to restaurants with higher transaction volumes.

\csection{Concluding Remark}

This paper proposes a novel QSAR model which characterizes peer effects via a quantile-based social norm. The quantile level is unknown and estimated from the data to identify the influential peer segment. We construct a GMM framework with pseudo-response instruments and kernel smoothing for estimation and statistical inference, establish equilibrium uniqueness and model identification, and derive the asymptotic distributions of our estimators.

Several potential topics can be considered for future research. First, extending the QSAR model to incorporate multiple quantile social norms would allow the joint influence of different peer segments to be examined.
Second, generalizing the QSAR model to accommodate nonlinear peer and covariate effects could broaden its applicability.
We believe that these efforts would expand the usefulness of our proposed QSAR model.

\begin{sidewaystable}[p]
\centering
\begin{threeparttable}
\caption{Performance Metrics Across Different Network Types and Node Sizes
(True Quantile Level \(z_0=0.5\)).}
\label{tab:network_metrics_0.5_new}

\setlength{\tabcolsep}{2.5pt}

\begin{tabular*}{\textwidth}{
@{\extracolsep{\fill}}
ll
*{15}{S[table-format=1.3]}
@{}
}
\toprule
\textbf{Network}
& \textbf{\(\boldsymbol\theta\)}
& \multicolumn{5}{c}{\textbf{\(n=1000\)}}
& \multicolumn{5}{c}{\textbf{\(n=2000\)}}
& \multicolumn{5}{c}{\textbf{\(n=3000\)}} \\
\cmidrule(lr){3-7}
\cmidrule(lr){8-12}
\cmidrule(lr){13-17}
&
& {\textbf{RMSE}} & {\textbf{Bias}} & {\textbf{SE}}
& {\textbf{tSE}} & {\textbf{CP}}
& {\textbf{RMSE}} & {\textbf{Bias}} & {\textbf{SE}}
& {\textbf{tSE}} & {\textbf{CP}}
& {\textbf{RMSE}} & {\textbf{Bias}} & {\textbf{SE}}
& {\textbf{tSE}} & {\textbf{CP}} \\
\midrule
SBM
& \(\beta_1\)
& 0.012 & 0.001 & 0.012 & 0.012 & 0.946
& 0.008 & 0.000 & 0.008 & 0.008 & 0.962
& 0.006 & 0.000 & 0.006 & 0.006 & 0.966 \\
& \(\beta_2\)
& 0.010 & 0.000 & 0.010 & 0.010 & 0.950
& 0.007 & 0.000 & 0.007 & 0.007 & 0.946
& 0.006 & 0.000 & 0.006 & 0.006 & 0.960 \\
& \(\lambda\)
& 0.049 & 0.006 & 0.048 & 0.051 & 0.966
& 0.038 & 0.005 & 0.038 & 0.041 & 0.968
& 0.033 & 0.004 & 0.033 & 0.036 & 0.970 \\
& \(z\)
& 0.018 & 0.000 & 0.018 & 0.019 & 0.946
& 0.011 & 0.000 & 0.011 & 0.012 & 0.966
& 0.009 & 0.000 & 0.009 & 0.009 & 0.952 \\

\addlinespace[0.5em]

PowerLaw
& \(\beta_1\)
& 0.012 & 0.000 & 0.012 & 0.012 & 0.934
& 0.008 & 0.001 & 0.008 & 0.008 & 0.952
& 0.007 & 0.000 & 0.007 & 0.007 & 0.954 \\
& \(\beta_2\)
& 0.010 & 0.000 & 0.010 & 0.010 & 0.942
& 0.007 & 0.000 & 0.007 & 0.007 & 0.958
& 0.005 & 0.000 & 0.005 & 0.006 & 0.966 \\
& \(\lambda\)
& 0.052 & 0.005 & 0.052 & 0.054 & 0.960
& 0.040 & 0.006 & 0.040 & 0.043 & 0.956
& 0.038 & 0.006 & 0.037 & 0.038 & 0.948 \\
& \(z\)
& 0.024 & 0.001 & 0.024 & 0.021 & 0.952
& 0.014 & 0.001 & 0.014 & 0.013 & 0.946
& 0.011 & 0.000 & 0.011 & 0.010 & 0.942 \\

\addlinespace[0.5em]

Dyad
& \(\beta_1\)
& 0.012 & 0.001 & 0.012 & 0.012 & 0.950
& 0.008 & 0.000 & 0.008 & 0.008 & 0.948
& 0.007 & 0.001 & 0.007 & 0.007 & 0.940 \\
& \(\beta_2\)
& 0.009 & 0.000 & 0.009 & 0.010 & 0.948
& 0.007 & 0.000 & 0.007 & 0.007 & 0.936
& 0.006 & 0.000 & 0.006 & 0.006 & 0.942 \\
& \(\lambda\)
& 0.061 & 0.007 & 0.060 & 0.060 & 0.946
& 0.043 & 0.008 & 0.042 & 0.047 & 0.976
& 0.038 & 0.008 & 0.037 & 0.041 & 0.960 \\
& \(z\)
& 0.022 & 0.001 & 0.022 & 0.020 & 0.968
& 0.012 & 0.000 & 0.012 & 0.012 & 0.962
& 0.008 & 0.000 & 0.008 & 0.008 & 0.950 \\
\bottomrule
\end{tabular*}
\begin{tablenotes}[flushleft]
\footnotesize
\item[] Notes: tSE denotes the average theoretical standard error computed from the estimated variance; CP denotes the empirical coverage of the 95\% Wald interval; and SBM denotes the stochastic block model. Results are based on \(R=500\) replications.
\end{tablenotes}
\end{threeparttable}
\end{sidewaystable}

\begin{table}[!htbp]
\centering
\begin{threeparttable}
\caption{Estimation Results for the Restaurant Data under LIM, CES, and QSAR.}
\label{tb:res}
\small
\setlength{\tabcolsep}{0pt}
\begin{tabular*}{\textwidth}{
@{}
>{\raggedright\arraybackslash}p{0.34\textwidth}
*{3}{
S[
  table-format=-1.3,
  table-space-text-post=***,
  table-column-width=0.135\textwidth
]
S[
  table-format=1.3,
  table-column-width=0.085\textwidth
]
}
@{}
}
\toprule
& \multicolumn{2}{c}{LIM}
& \multicolumn{2}{c}{CES}
& \multicolumn{2}{c}{QSAR} \\
Variable
& {Estimate} & {SE}
& {Estimate} & {SE}
& {Estimate} & {SE} \\
\midrule
Repeat Customer Ratio     & 0.318***  & 0.016 & 0.318***  & 0.016 & 0.317***  & 0.017 \\
Commercial Zone Diversity & 0.100***  & 0.013 & 0.102***  & 0.012 & 0.102***  & 0.013 \\
Menu Variety              & 0.025*    & 0.014 & 0.024*    & 0.014 & 0.027*    & 0.014 \\
Average Transaction Value & 0.175***  & 0.016 & 0.181***  & 0.015 & 0.187***  & 0.015 \\
Operating Hours           & 0.058**   & 0.028 & 0.056**   & 0.028 & 0.053*    & 0.028 \\
Network Coefficient       & -0.689*** & 0.075 & -0.623*** & 0.094 & -0.618*** & 0.068 \\
\addlinespace[0.25em]
CES elasticity parameter / QSAR quantile level  & {---}     & {---} & 0.342     & 0.734 & 0.427     & 0.063 \\
\bottomrule
\end{tabular*}
\vspace{0.25em}
\begin{tablenotes}[flushleft]
\footnotesize
\item[] Notes: Estimation results and standard errors are reported. All models include an intercept, which is omitted. Statistical significance is indicated by ${}^{***}p<0.01$, ${}^{**}p<0.05$, and ${}^{*}p<0.10$. No significance markers are reported for the CES elasticity parameter and QSAR quantile level.
\end{tablenotes}
\end{threeparttable}
\end{table}

\begin{table}[!htbp]
\centering
\begin{threeparttable}
\caption{QSAR Estimates Across Different Groups (Low/High).}
\label{tab:qsar_two_groups}
\small
\setlength{\tabcolsep}{0pt}
\begin{tabular*}{\textwidth}{
@{}
>{\raggedright\arraybackslash}p{0.34\textwidth}
*{2}{
S[
  table-format=-1.3,
  table-space-text-post=***,
  table-column-width=0.20\textwidth
]
S[
  table-format=1.3,
  table-column-width=0.13\textwidth
]
}
@{}
}
\toprule
& \multicolumn{2}{c}{Low}
& \multicolumn{2}{c}{High} \\
Variable
& {Estimate} & {SE}
& {Estimate} & {SE} \\
\midrule
Repeat Customer Ratio     & 0.246***  & 0.045 & 0.345***  & 0.023 \\
Commercial Zone Diversity & 0.072***  & 0.022 & 0.119***  & 0.020 \\
Menu Variety              & 0.046**   & 0.022 & 0.002     & 0.017 \\
Average Transaction Value & 0.187***  & 0.033 & 0.178***  & 0.018 \\
Operating Hours           & 0.018     & 0.042 & 0.087**   & 0.037 \\
Network Coefficient       & -0.609*** & 0.098 & -0.685*** & 0.101 \\
\addlinespace[0.25em]
QSAR quantile level              & 0.360     & 0.067 & 0.577     & 0.096 \\
\bottomrule
\end{tabular*}
\vspace{0.25em}
\begin{tablenotes}[flushleft]
\footnotesize
\item[] Notes: Estimation results and standard errors are reported. All models include an intercept, which is omitted. Statistical significance is indicated by ${}^{***}p<0.01$, ${}^{**}p<0.05$, and ${}^{*}p<0.10$. No significance markers are reported for the QSAR quantile levels.
\end{tablenotes}
\end{threeparttable}
\end{table}

\bibliography{reference}

\end{document}